\documentclass[aps,preprint,12pt,floatfix,nofootinbib]{revtex4-2}

\usepackage{amsmath}
\usepackage{amssymb}
\usepackage{xcolor}
\usepackage{graphicx}
\usepackage{dcolumn}
\usepackage{float}
\usepackage{bm}
\usepackage{caption}
\usepackage{subcaption}
\usepackage{epstopdf}
\usepackage{epsfig}
\usepackage{adjustbox}
\usepackage{times}
\usepackage{natbib}
\usepackage{pdfrender}
\usepackage{natbib}
\bibpunct{(}{)}{;}{a}{}{,}

\usepackage[italicdiff]{physics}
\usepackage{hyperref}

\hypersetup{colorlinks=true,
            allcolors=blue}

\newcommand{\orcidicon}[1]{\href{https://orcid.org/#1}{\includegraphics[height=\fontcharht\font`\B]{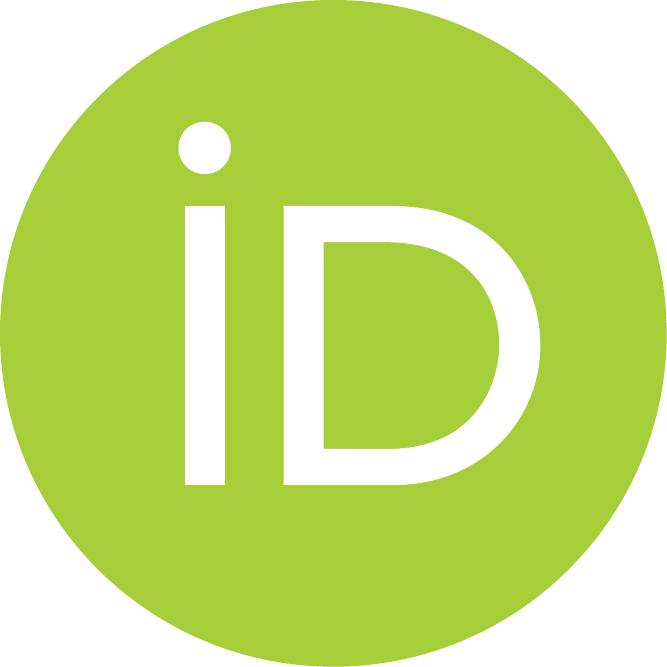}}}

\newcommand\blfootnote[1]{%
  \begingroup
  \renewcommand\thefootnote{}\footnote{#1}%
  \addtocounter{footnote}{-1}%
  \endgroup
}

\newcommand*{\boldcheckmark}{%
  \textpdfrender{
    TextRenderingMode=FillStroke,
    LineWidth=.5pt, 
  }{\checkmark}%
  }

\begin{document}

\title{Constraining the parameterized neutron star equation of state with astronomical observations}

\author{Jaikhomba Singha$^{1}$\,\orcidicon{0000-0002-1636-9414}}
\author{S. Mullai Vaneshwar$^2$}
\author{Ankit Kumar$^1$\,\orcidicon{0000-0003-3639-6468}}

\affiliation{$^\mathit{1}$Department of Physics, Indian Institute of Technology Roorkee, Roorkee 247667, India}
\affiliation{$^\mathit{2}$Department of Physics, National Institute of Technology Calicut, Kozhikode 673601, India}

\blfootnote{Jaikhomba Singha: \textcolor{blue}{ mjaikhomba@gmail.com}}

\begin{abstract}
We utilise the phenomenologically parameterized piecewise polytropic equations of state to study various neutron star properties. We investigate the compliance of these equations of state with several astronomical observations. We also demonstrate that the theoretical estimates of the fractional moment of inertia cannot explain all the pulsar glitches observed. We model the crust as a solid spheroidal shell to calculate the fractional moment of inertia of fast-spinning neutron stars. We also show that the braking index obtained in a simple magnetic dipole radiation model with a varying moment of inertia deviates significantly from the observed data. Future developments in both theory and observations may allow us to use the fractional moment of inertia and braking index as observational constraints for neutron star equation of state.
\end{abstract}

\maketitle

\section{Introduction}           
\label{sect:intro}

Neutron stars (NSs) are extremely dense objects formed as a result of violent supernovae explosions of massive stars at the end of their lifetimes. Their extreme properties, such as strong magnetic fields, stable rotations, intense gravity, etc., make them ideal laboratories to test various theories of physics under extreme conditions \citep{ ns0, ns00,ns1, ns2, ns3}. The observational manifestation of a NS is a pulsar. The internal structure and composition of NSs are governed by the equation of state (EOS) of neutron-rich matter \citep{Lattimer_2001}. Despite extensive research, the composition of NS matter is not precisely known. While the primary constituents are neutrons and protons, the possible existence of hyperons and kaon condensates is still being debated. It is understood that the observation of most massive NSs could reveal some information about the presence of such exotic matter. A detailed study of EOS over a wide range of densities is crucial in understanding the properties of NSs. NSs are also continuously losing their rotational energy due to the emission of several high energetic particles. This loss is reflected in the spin evolution of pulsars. The study of the spin evolution of pulsars helps in understanding the interior and the exterior of NSs \citep{spin0, spin1, spin2}.

In this work, we utilise the piecewise polytropic EOSs to study the various important properties of NSs. Several pieces of the relativistic polytropes are ensured to be thermodynamically continuous while mimicking various phase transitions at high densities. We further put various constraints on these EOSs obtained from observations. We show that the fractional moment of inertia (FMI) obtained from theory cannot explain all the observed glitches. We also investigate how FMI varies with rotational frequencies. Finally, we demonstrate the deviation of the observed braking indices from its theoretical estimates.

The paper has been organised in the following manner. In Sec. \ref{formalism}, we briefly describe the structure of non-rotating and rotating neutron stars and the ways to estimate several NS properties like mass, radius, tidal deformability, fractional moment of inertia and braking index. In Sec. \ref{results}, we present the results obtained for the various EOSs used in the paper and discuss their implications. The several observational measurements used as constraints in this work have been listed in this section. Finally, in Sec. \ref{conclusion}, we end the paper by presenting the conclusions of this work.


%

\section{Formalism}
\label{formalism}
The equilibrium configurations of NSs are usually calculated in two steps. The EOS of high-density matter is estimated at first, which is thereafter utilised for NS structure calculations in accordance with the principles of general relativity. In this work, we have constructed a few well-known EOSs based the on piecewise polytropic formalism (in \cite{Read2009, LackeyThesis}) for mimicking various phase transitions at high densities.


\subsection{Structure of non-rotating NSs}
The structure of static and spherical NSs are governed by the following Tolman-Oppenheimer-Volkov (TOV) equations \citep{TOV1,TOV2}:
\begin{equation}
\dv{r}p(r) = -\displaystyle\frac{\epsilon(r)M(r)}{r^2} \qty[ 1 + \frac{p(r)}{\epsilon(r)} ] \qty[ 1 + \frac{4\pi{r}^3p(r)}{M(r)} ] \qty[ 1 - \frac{2M(r)}{r} ]^{-1},
\end{equation}
\begin{equation}
\dv{r}\nu(r) = \frac{1}{\epsilon(r) + p(r)} \displaystyle\frac{dp(r)}{dr},
\end{equation}
\begin{equation}
\dv{r}M(r) = 4\pi r^2\epsilon(r),
\end{equation}
where \begin{math}M(r)\end{math} is the enclosed mass of the NS within a radius $r$, and $\nu(r)$ is the metric potential.

\subsection{Fractional Moment of Inertia of slow rotating NSs}
\label{sec:2.3}
Assuming the NS to be rotating slowly, the moment of inertia (MOI) can be calculated within the Hartle-Thorne's approximation \citep{Hartle, HartleThorne}. Under this condition, the metric takes the following form:
\begin{equation}
    ds^2 = -e^{2\nu(r)}dt^2 + \qty( 1 - \frac{2M(r)}{r} )^{-1}dr^2  - 2\omega(r) \ r^2\sin^2{\theta} \ d\phi \ dt + r^2 \ d\theta^2 + r^2\sin^2{\theta} \ d\phi^2,
\end{equation}
where $\omega(r)$ is the frame dragging frequency. The NSs assume a near-spherical shape, and the MOI can be obtained as
\begin{equation}
    I = \frac{8\pi}{3} \int_{0}^{R} dr \ r^4 \frac{\epsilon(r) + p(r)}{\qty[ 1-2M(r)/r ]} \qty( 1-\frac{\omega(r)}{\Omega} ) e^{-\nu (r)},
\end{equation}
where $\Omega$ is the spin frequency of the NS. The crustal MOI has the same form
\begin{equation}
    \Delta I = \frac{8\pi}{3} \int_{R_c}^{R} dr \ r^4 \frac{\epsilon(r) + p(r)}{\qty[ 1-2M(r)/r ]} \qty( 1-\frac{\omega(r)}{\Omega} ) e^{-\nu (r)}
\end{equation}
where $R_c$ is the core-crust transition radius. The FMI is now defined as the ratio $\Delta I/I$.

\subsection{Tidal Deformability}

A NS experiences a tidal gravitational field in the presence of a companion. The tidal deformability parameter is defined as \citep{Hinderer2010}
\begin{eqnarray}
    \lambda = -\frac{Q_{ij}}{\mathcal{E}_{ij}},
\end{eqnarray}
where $Q_{ij}$ is the induced quadrupole moment, due to the tidal field $\mathcal{E}_{ij}$. The tidal deformability can be expressed in terms of the love number $k_2$ and the NS radius $R$ as
\begin{equation}
    \lambda = \frac{2}{3}k_2 R^5.
\end{equation}
The love number $k_2$ is given by
\begin{eqnarray}
    k_2 = \frac{8C^5}{5}(1-2C)^2 \qty[ 2 + 2C(y_R -1) - y_R ] \times  \Big\{ 2C(6-3y_R + 3C(5y_R -8)) \nonumber \\ +4C^3 \qty[ 13-11y_R+C(3y_R-2)+2C^2(1+y_R) ] \\
    + 3(1-2C)^2 \qty[ 2-y_R+2C(y_R-1) ] \log(1-2C)  \Big\} ^{-1}, \nonumber
\end{eqnarray}
where $C = M/R$ is the compactness parameter, and $y_R$ satisfies
\begin{equation}
    r\dv{r}y(r) + y(r)^2 + y(r)F(r) + r^2Q(r) = 0,
\end{equation}
with
\begin{equation}
F(r) = \frac{r - 4\pi r^3 \qty[ \epsilon - p(r) ]}{r-2M(r)}, \ \text{and}
\end{equation}
\begin{equation}
Q(r) = \frac{4\pi r \qty[ 5\epsilon(r) + 9p(r) + \frac{\epsilon(r) + p(r)}{\delta p(r)/\delta\epsilon(r)} - \frac{6}{4 \pi r^2} ]}{r-2M(r)} - 4 \Bigg\{ \frac{M(r) + 4\pi r^3p(r)}{r^2 \qty[ 1-2M(r)/r ]}\Bigg \}^2.
\end{equation}
The dimensionless tidal deformability can be defined as
\begin{equation}
    \Lambda = \frac{2}{3}k_2 C^{-5}.
\end{equation}

\subsection{Pulsar Braking Index}
NSs emit electromagnetic energy via magnetic dipole radiation. This comes at the expense of the rotational kinetic energy, and hence the NSs are spinning down constantly with time. The pulsar braking index is related to the spin frequency as
\begin{equation}
    n(\Omega) = \frac{\Omega \ddot{\Omega}}{\dot{\Omega}^2}.
\end{equation}
In terms of spin period ($P$) and its derivatives,
\begin{equation}
    n(P) = 2 - \frac{P \ddot{P}}{\dot{P}^2}.
\end{equation}
If we consider the MOI to be frequency independent,  $n=3$ \citep{Kaspi1994}. However, the MOI varies with the spin frequency, and hence it also varies with time \citep{Glendenning1997}. Assuming the MOI to be frequency dependent we have \citep{Hamil2015}
\begin{equation}
    n (\Omega) = 3 - \frac{3 \Omega I' + \Omega^2 I''}{2I + \Omega I'}.
    \label{bi}
\end{equation}

\subsection{Fractional Moment of Inertia of fast spinning NSs}
\label{sec:2.6}
For the special case of fast spinning NSs we calculate the equilibrium NS configurations in an axially symmetric space-time. In this case the infinitesimal line element is given by
\begin{equation}
    ds^2 = -N^2dt^2 + A^2(dr^2+r^2d\theta^2) + B^2r^2\sin^2\theta(d\phi-N^\phi dt)^2,
\end{equation}
where $N$, $A$, $B$, and $N^\phi$ are the metric functions dependent on $r$ and $\theta$. The numerical computations for solving the Einstein field equations are performed with suitable adaption of \texttt{LORENE} libraries\footnote{\url{https://lorene.obspm.fr/}} \citep{gourgoulhon2011introduction}. The NS assumes a spheroidal shape, and the moment of inertia (MOI) for a rotation frequency $\Omega$ is given by
\begin{equation}
I = \frac{2}{\Omega} \int_{0}^{\pi/2}d\theta \int_{0}^{R(\theta)} dr \ A^2(r,\theta)B^2(r,\theta)\qty[\epsilon(r,\theta)+p(r,\theta)]\frac{U(r,\theta)}{1-U^2(r,\theta)}r^3\sin^2\theta,
\end{equation}
and $R(\theta)$ is the NS radius in the $\theta$ direction, and $U$ is defined as
\begin{equation}
    U(r,\theta) = \frac{B(r,\theta)}{N(r,\theta)} \qty[ \Omega-N^\phi (r,\theta) ]r \sin{\theta}.
\end{equation}
The baryon density profile is $\theta$ dependent, and so must be the core-crust transition radius $R_c$. We can therefore write the crustal MOI as
\begin{equation}
\Delta I = \frac{2}{\Omega} \int_{0}^{\pi/2}d\theta \int_{R_c(\theta)}^{R(\theta)} dr \ A^2(r,\theta)B^2(r,\theta)\qty[\epsilon(r,\theta)+p(r,\theta)]\frac{U(r,\theta)}{1-U^2(r,\theta)}r^3\sin^2\theta.
\end{equation}
$\Delta I/I$ is defined as the FMI.

\section{Results and discussions}
\label{results}

The NS interior is broadly divided into two major regions, i.e., the crust and the core. The crust is primarily composed of nucleons, and is well understood because of its near-nuclear density. The core is highly compressed to many times the nuclear density, and despite extensive research, its composition and interactions are not precisely known. It is expected that the high-density matter in the NS core may show multiple phase transitions due to the sequential onset of exotic particles.
\begin{figure}[b]
\centerline{\includegraphics[scale=0.55]{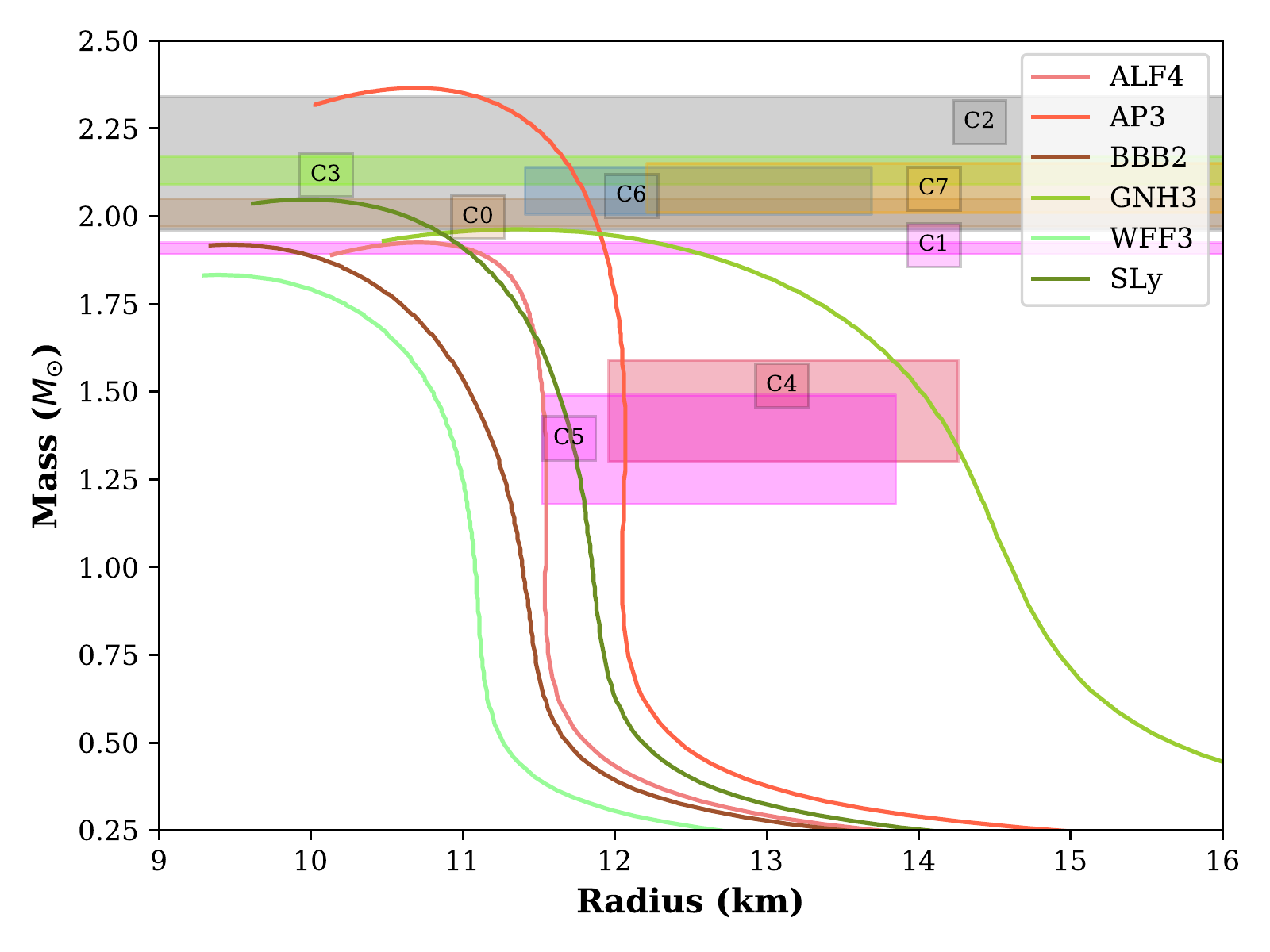}}
\caption{The mass-radius relations for the ALF4, AP3, BBB2, GNH3, WFF3 and SLy EOSs. The shaded regions depict the imposed mass-radius constraints from astronomical observations (see Table \ref{table:mr}).}
\label{fig:MR}
\end{figure}
\begin{table}[H]
\centering
{\begin{tabular}{llclc}
\hline
\hline
PSR- J Name & $M$ ($\textup{M}_\odot$) & $R$ (km) & Reference & Constraint\\
\hline
\hline
\centering
J0348+0432 & 2.01$\pm$0.04 & - & \cite{Antoniadis_2013} & C0\\
J1614-2230 & 1.908$\pm$0.016 & - & \cite{Demorest2010, Arzoumanian_2018} & C1\\
J0740+6620 & 2.14$^{+0.20}_{-0.18}$ & - & \cite{Cromartie2020} & C2\\
J1810+1744 & 2.13$\pm$0.04 & - & \cite{Romani_2021} & C3\\
\hline
J0030+0451 & 1.44$^{+0.15}_{-0.14}$ & 13.02$^{+1.24}_{-1.06}$ & \cite{Miller_2019} & C4\\
 & 1.34$^{+0.15}_{-0.16}$ & 12.71$^{+1.14}_{-1.19}$ & \cite{Riley_2019} & C5\\
 \hline
J0740+6620 & 2.072$^{+0.067}_{-0.066}$ &  12.39$^{+1.30}_{-0.98}$ & \cite{Riley_2021} & C6\\
 & 2.08$\pm$0.07 &  13.71$^{+2.67}_{-1.50}$ & \cite{Miller_2021} & C7\\
\hline
\hline
\end{tabular}}
\caption{Neutron star mass and radius constraints from various astronomical observations. $M$ is the mass in units of the solar mass $M_\odot$, and $R$ is the NS radius.}
\label{table:mr}
\end{table}

Following the seminal approach of \cite{Read2009, LackeyThesis}, we have utilised the low density SLy EOS for the crust \citep{SLy}, and the core is modeled with a variety of piecewise polytropes to emulate the expected phase transitions at high densities. This includes the ALF4 \citep{ALF4}, AP3 \citep{AP3}, BBB2 \citep{BBB2}, GNH3 \citep{GNH3}, WFF3 \citep{WFF3} and SLy \citep{SLy} EOSs. ALF4 is a hybrid EOS with mixed APR nuclear matter and colour-flavor locked quark matter EOS, AP3, and WFF3 are based on the variational method, BBB2 is a non-relativistic EOS, and SLy is a potential-method EOS, with all of them modeling the $npe\mu$ matter. On the other hand, GNH3 is based on the relativistic mean field theory, and takes into account the contribution of hyperons as well. The polytropic parameters of all the EOSs used in this work are taken from \cite{Read2009, LackeyThesis}.

\begin{table}
\centering
{\begin{tabular}{clc}
\hline
\hline
$\Lambda_{1.4}$ & Reference & Constraint\\
\hline
\centering
$70 - 580$ & \cite{Abbott_2018} & C8\\
$240 - 730$ & \cite{Jiang_2020} & C9\\
$133 - 686$ & \cite{Li2021} & C10\\
\hline
\hline
\end{tabular}}
\caption{Gravitational wave constraints on the dimensionless tidal deformability $\Lambda$. $\Lambda_{1.4}$ is the dimensionless tidal deformability for a 1.4 solar mass neutron star.}
\label{table:tid}
\end{table}

There have been several efforts to constraint the NS EOS with astronomical observations and simulations. While the most common constraint is the measurement of the mass (see C0-C3 in Table \ref{table:mr}), there has been a good progress on the much awaited simultaneous estimation of the radius (see C4-C7 in Table \ref{table:mr}). These serve as stringent limits on the allowed mass-radius configurations of stationary NSs. Fig. \ref{fig:MR} shows the mass-radius curves for various EOSs with the shaded regions depicting the various mass-radius constraints. WFF3 does not satisfy any, ALF4, BB2, SLy, and GNH3 satisfy only some, and AP3 satisfies most of the mass-radius constraints. The inclusion of hyperons in GNH3 makes it a soft EOS, and this shows up in the comparatively large radii of the allowed configurations. The mass and radius obtained for the various EOSs match the results given in \cite{Read2009} and \cite{LackeyThesis}. Any valid EOS must explain the existence of the most massive neutron star PSR J0348+0432, and hence C0 is a necessary constraint to comply with. It can be seen that the maximum allowed mass with ALF4, BBB2, and WFF3, is not high enough to satisfy the C0 constraint, hence we no longer include these EOSs in further discussions.

\begin{figure}
\centerline{\includegraphics[scale=0.55]{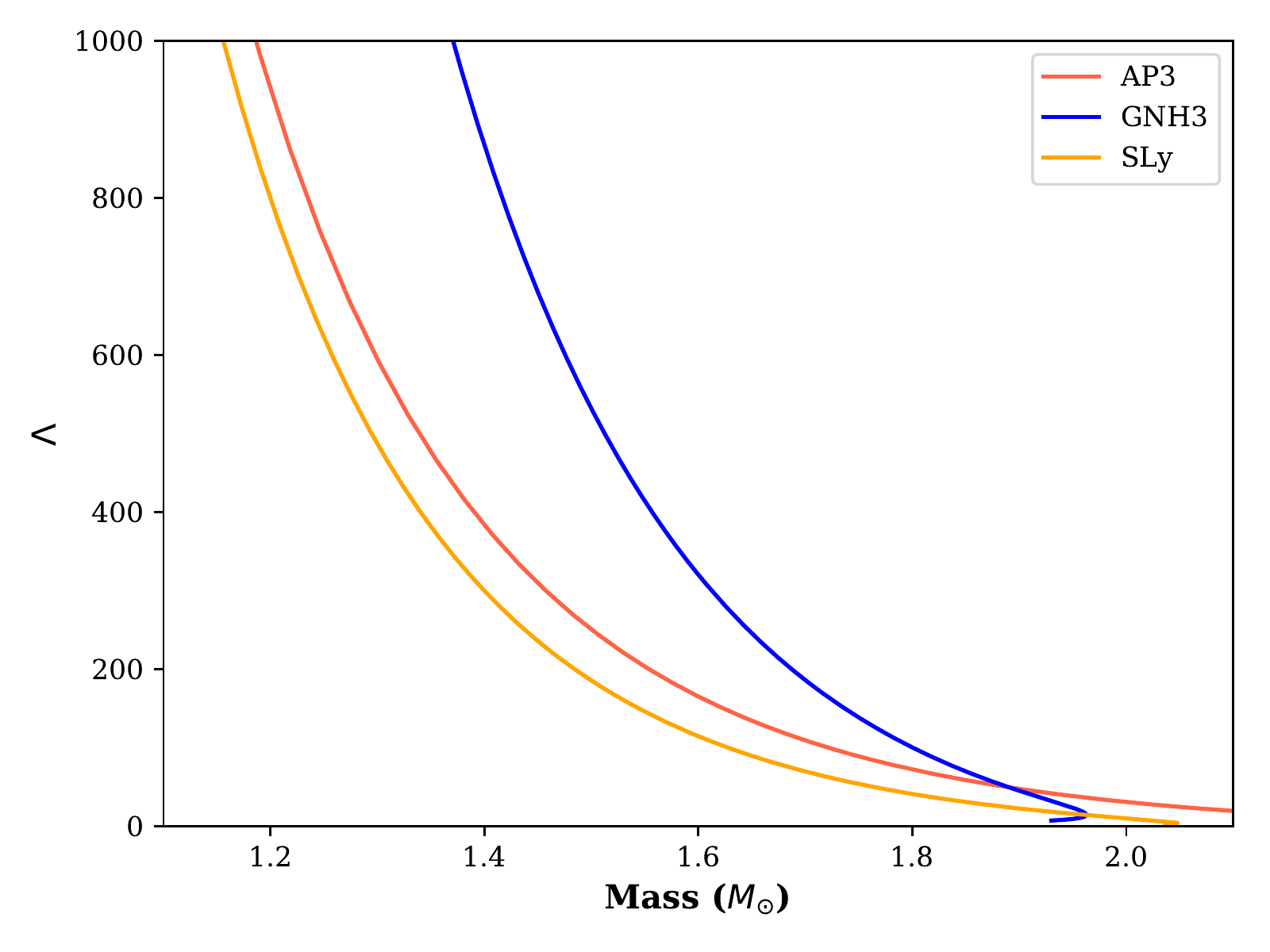}}
\caption{The dimensionless tidal deformability, $\Lambda$ as a function of NS mass for several EOSs.}
\label{lambda}
\end{figure}

LIGO's paradigmatic observation of the merger of two neutron stars (GW170817) quantified the response of NS matter towards a strong tidal gravitational field. It has been shown by \cite{TD4, TD2, TD1, TD3, TD5}; and many others that the determination of tidal deformability has been crucial in constraining the high-density EOS. The tidal deformability constraints used in this work are tabulated in Table \ref{table:tid}. In Fig. \ref{lambda}, we plot $\Lambda$ as a function of the NS mass. GNH3 satisfies none, and AP3 and Sly satisfy all of the three deformability constraints. As expected from previous discussions, the NSs described by the soft GNH3 EOS will have smaller compactness, and hence a larger $\Lambda$. The compliance of all the EOSs with the imposed constraints is summarised in Table \ref{table_constraints}. It can be seen that AP3 satisfies the maximum number of constraints. It explains the existence of the massive pulsar PSR J0348+0432~\citep{Antoniadis_2013}.

\begin{table}
\centering
\begin{adjustbox}{width=0.75\textwidth}
{\begin{tabular}{c||ccc|cccc||ccc}
\hline
\hline
\centering
 &\multicolumn{10}{c}{Constraint}\\
 \cline{2-11}
& \multicolumn{3}{c}{Mass} & \multicolumn{4}{c}{Mass-Radius} & \multicolumn{3}{c}{Tidal Deformability} \\
EOS  & C1 & C2 & C3 &C4& C5& C6& C7& C8& C9& C10\\
\hline
\centering
AP3   & $\boldcheckmark$ & $\boldcheckmark$ & $\boldcheckmark$ & $\boldcheckmark$ & $\boldcheckmark$ & $\boldcheckmark$ & - & $\boldcheckmark$ & $\boldcheckmark$ & $\boldcheckmark$ \\
GNH3  & $\boldcheckmark$ & - & - & $\boldcheckmark$ & - & - & - & - & - & - \\
SLy  & $\boldcheckmark$ & - & - & - & $\boldcheckmark$ & - & - & $\boldcheckmark$ & $\boldcheckmark$ &  $\boldcheckmark$\\
\hline
\hline
\end{tabular}}
\end{adjustbox}
\caption{Compliance of various EOSs with the imposed mass (C1-C3), mass-radius (C4-C7), and tidal deformability constraints (C8-C10). The checkmark indicates that the constraint is satisfied.}
\label{table_constraints}
\end{table}
\begin{figure}[b]
\centerline{\includegraphics[scale=0.75]{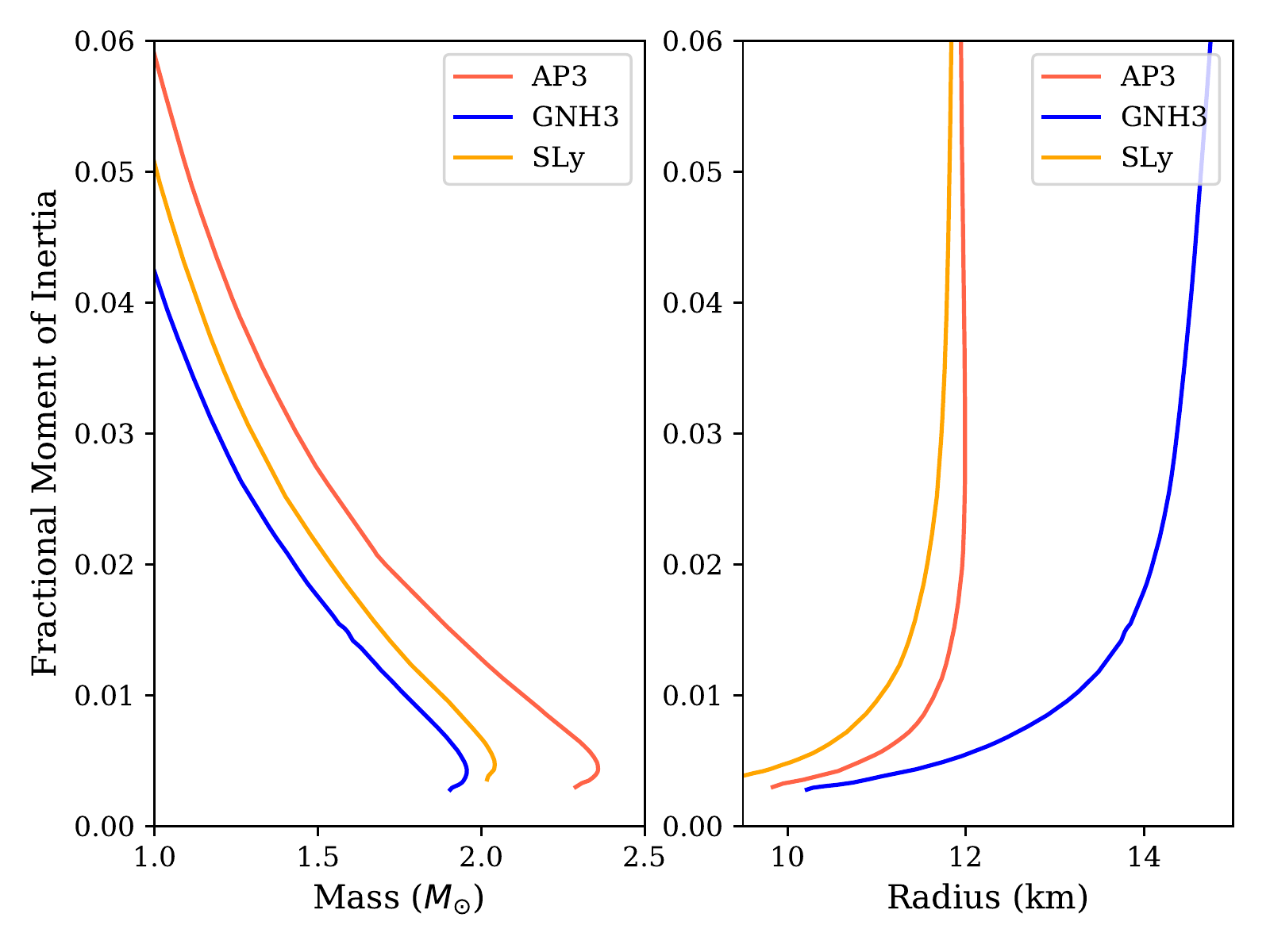}}
\caption{The fractional moment of inertia as a function of NS masses (left) and radii (right) for the several piecewise polytropic EOS resembling the realistic EOSs: AP3, GNH3, and SLy.}
\label{fig:fmi_r}
\end{figure}
We now discuss some other properties of neutron stars viz., fractional moment of inertia (FMI) and braking index. There are times when a NS suddenly spins up. Such an event is called a pulsar glitch \citep{RADHAKRISHNAN1969}, and their observations help in various ways to probe the interior of NSs \citep{Gl1, Gl2, Gl3}. Presently, the most favourable model to explain the occurrence of a pulsar glitch is based on pinning of superfluid vortices in the NS crust \citep{haskell_melatos}. It is believed that glitches occur due to the transfer of angular momentum from the interior to the outer crust.  The ratio of the MOI of the crust and the core, i.e. FMI, can be estimated from observations of pulsar glitches \citep{Basu_2018}
\begin{equation}
    \text{FMI} = \frac{\Delta I}{I} > 2 \tau_i \frac{1}{t_i} \left ( \frac{\Delta \nu}{\nu} \right ),
\end{equation}
where $\tau_i$ is the characteristic age of the pulsar, $t_i$ is the time preceding the last glitch, and $ \Delta \nu/\nu $ is the fractional rise in the spin frequency. As mentioned in Sec. \ref{sec:2.3} and \ref{sec:2.6}, FMI can also be estimated for a given EOS for both slow rotating and fast rotating NSs.
\begin{figure}[H]
\begin{minipage}{0.65\textwidth}
\includegraphics[scale=0.62, trim={0.6cm, 0.52cm, 0.2cm, 0.52cm}]{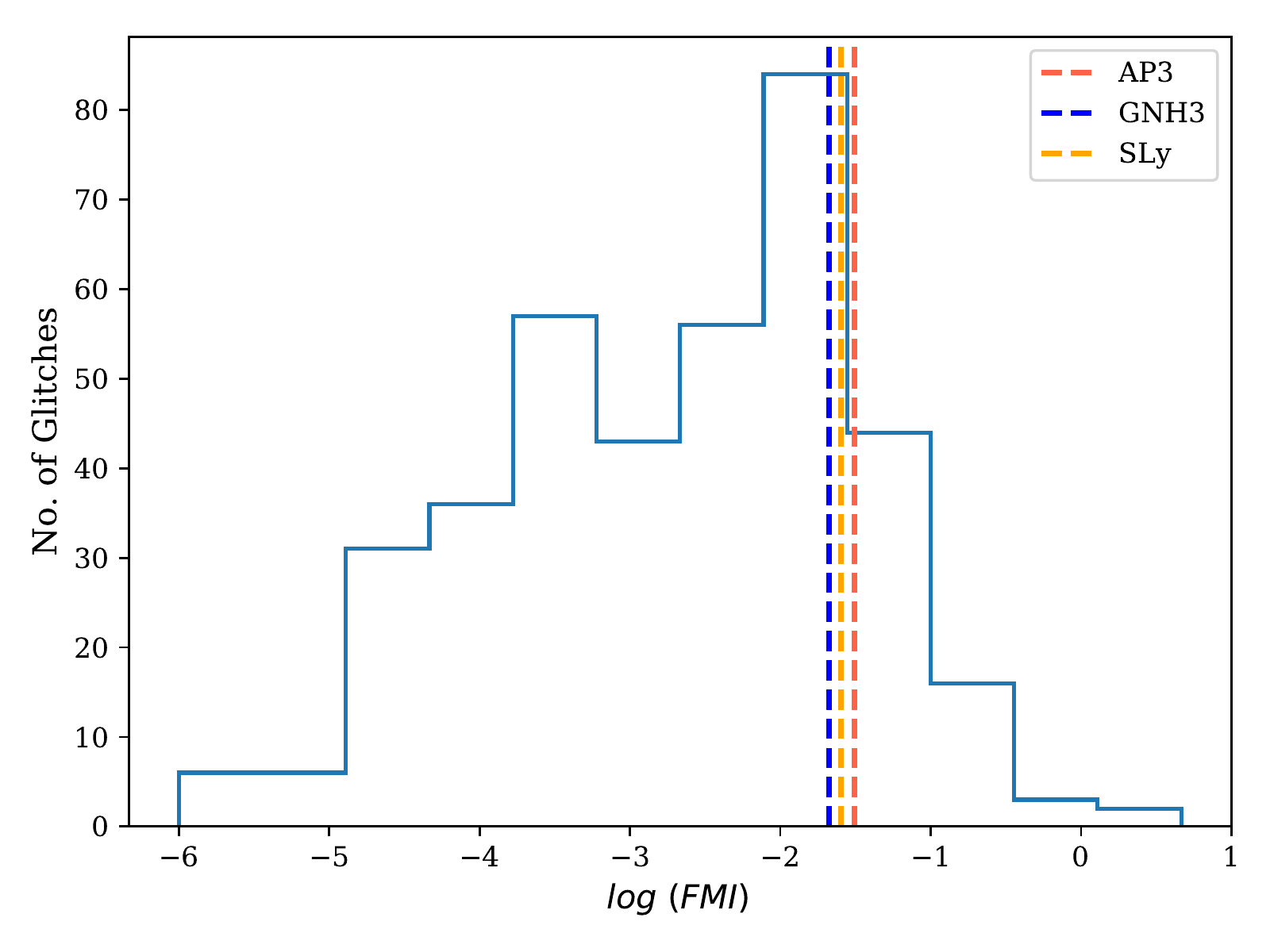}
\caption{Distribution of the FMI obtained from different glitch observations catalogued by the Jodrell Bank Observatory. The dotted lines depict the FMI obtained for the various EOSs for a canonical NS.}
\label{fig: fmi_dist}
\end{minipage}
\begin{minipage}{0.3\textwidth}
\captionsetup{type=table} 
\centering
\caption{The fraction of glitches which cannot be explaind with the estimated FMI from the EOSs for a canonical NS}.
{\begin{tabular}{cc}
\hline
\hline
EOS & $\bar{f} (\%)$\\
\hline
\centering
AP3 &  14.84\\
GNH3  & 21.88\\
SLy &  18.23\\
\hline
\hline
\end{tabular}}
\label{table:frac}
\end{minipage}
\end{figure}

Fig. \ref{fig:fmi_r} shows the fractional moment of inertia as a function of the NS mass and radius in the slow rotation limit. The crust-core transition density for every EOS ranges from $6.659 \times 10^{13}$ to $2.014 \times 10^{14}$ g/cc. It is seen that FMI is the lowest for GNH3. Since FMI can be deduced from observations, it can be utilised to constrain the NS as well. However, the FMI estimated for a particular EOS cannot explain all the glitches observed. Many studies suggest that in order to explain some of the glitches, the participation of the superfluidity in the core along with the crustal superfluid must be invoked \citep{Anderson, Basu_2018}. Although there has been some progress in theorizing superfluidity inside NS cores \citep{SFcore1, SFcore2}, we do not explore this in the present work. Fig. \ref{fig: fmi_dist} shows the distribution of FMIs of 384 glitches from the Jodrell Bank pulsar glitch catalogue\footnote{\url{https://www.jb.man.ac.uk/~pulsar/glitches/gTable.html}}. Since FMI cannot be estimated for the first glitch, we exclude the first glitch of every multiple glitching pulsars and also the pulsars with only one glitch. Assuming a canonical NS, for a fraction of glitches $\bar{f}$, the observed FMI is larger than what is estimated under the slow rotation approximation (see the dotted lines in Fig. \ref{fig: fmi_dist}). The values of $\bar{f}$ obtained for the various EOSs is summarised in Table \ref{table:frac}. It can again be seen that the AP3 explains the most number of glitches, while the GNH3 explains the least. For these glitches, it is expected that there is a contribution of superfluidity in the NS cores. It is important to point out that we have excluded the effect of entrainment. Presently, the FMI does not put very strong constraints on the EOS. Future advancements in theoretical studies along with better observational facilities may allow us to use FMI as a constraint for the NS EOS.

\begin{figure}
\centerline{\includegraphics[scale=0.8, trim={0.5cm, 0.52cm, 0.2cm, 0.52cm}]{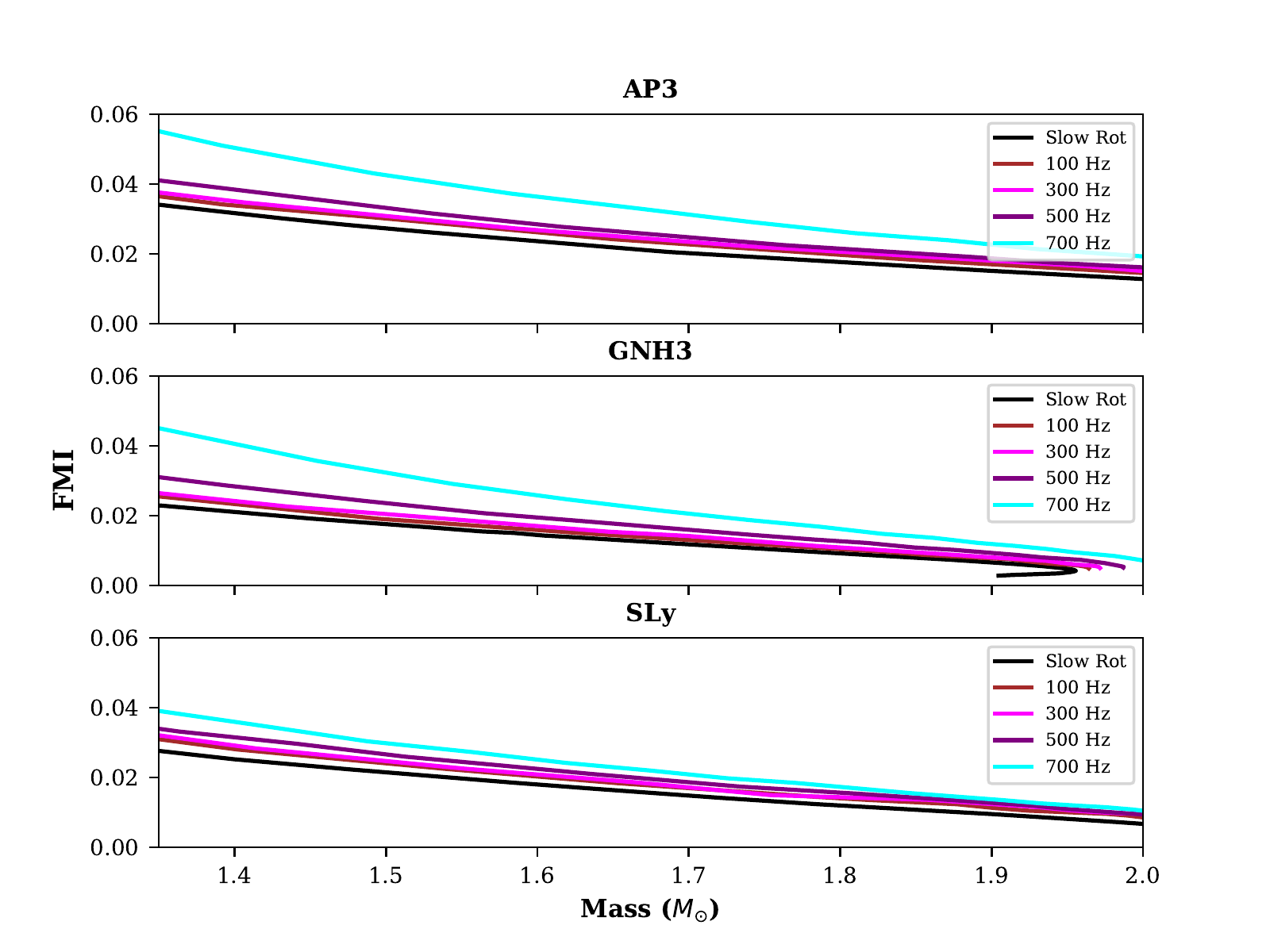}}
\caption{The fractional moment of inertia (FMI) as a function of NS mass for different rotation frequencies. $M_\odot$ is the solar mass, and the frequencies are mentioned in the legends. The calculations for the slowly rotating case are performed with in the Hartle-Thorne approximation (as discussed in Sec~\ref{sec:2.3}). The results for the fast rotating cases are calculated exactly with a suitable adaption of \texttt{LORENE} libraries (see Sec.\ref{sec:2.6}).}
\label{fig:fmi_rot}
\end{figure}

\begin{table}[b]
\centering
{\begin{tabular}{lccl}
\hline
\hline
PSR & $\Omega$ (rad/s) & Braking Index & Reference \\
\hline
\hline
\centering
B0531+21 (Crab) & 189.912 & 2.342 & \cite{Lyne2014}\\
B0833-45 (Vela) & 70.4 & 1.4 & \cite{Lyne1996}\\
B1509-58 & 41.680 & 2.839 & \cite{Livingstone2007}\\
J1846-0258 & 19.340 & 2.65 & \cite{Livingstone2007}\\
J1119-6127 & 15.401 & 2.684 & \cite{Weltevrede_2011}\\
\hline
\hline
\end{tabular}}
\caption{The braking index values obtained from several pulsar observations.}
\label{tab:br_tab}
\end{table}

In a first, we explored the variation of FMI at very high rotation rates. Fig. \ref{fig:fmi_rot} shows the FMI plotted as a function of NS masses at different frequencies. The FMI does not vary with the frequency in slow rotation approximation. We demonstrate a monotonic increment in the fractional moment of inertia with an increase in the spin frequency. We have limited our calculations up to $700$ Hz, which is close to the spin frequency of the fastest spinning radio pulsar observed at $716$ Hz \citep{716Hz}. At this frequency, Sly shows the least deviation. The centrifugal force on a spinning object is proportional to its distance from the rotation axis. The crust is the outermost region, and hence with an increase in spin frequency, $\Delta I$ changes relatively faster than $I$. This leads to an overall increment of the ratio $\Delta I/I$.

\begin{figure}
\centerline{\includegraphics[scale=0.55]{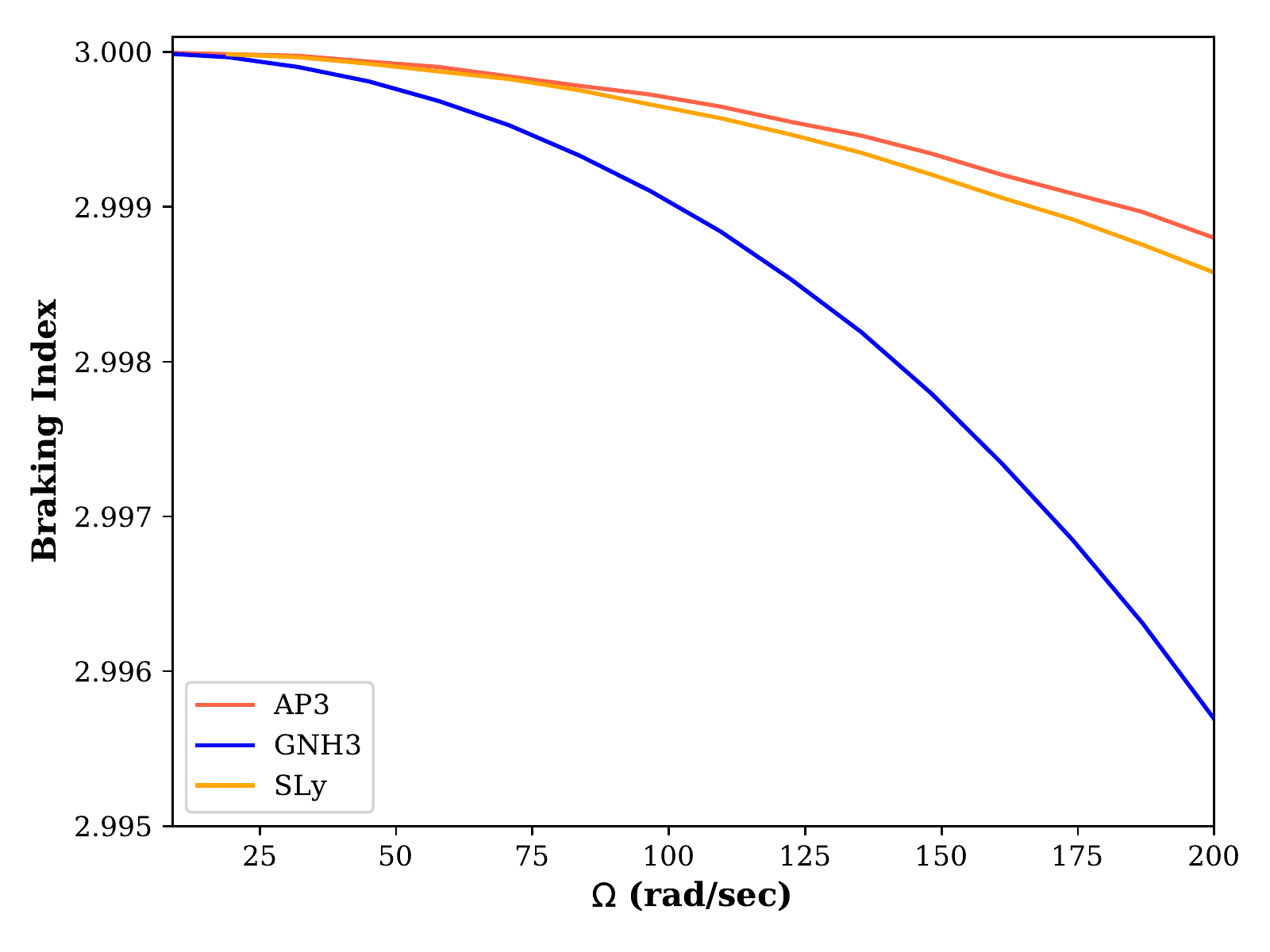}}
\caption{The braking index as a function of neutron star rotation frequency for the various EOSs.}
\label{fig:br_fig}
\end{figure}

Another quantity that we have explored is the pulsar braking index. NSs are constantly radiating electromagnetic energy via magnetic dipole radiation (MDR). This comes at the expense of the rotational kinetic energy, and hence the NSs are continuously slowing down. The braking index quantifies the rate of decrease of spin frequency. Considering the MOI to be frequency dependent, the braking index is given by Eq. \ref{bi}. The braking index of several pulsars have been estimated with the help of long baseline timing programs and other astronomical observations, and some of them are summarized in Table \ref{tab:br_tab}. It is evident that for many of these young pulsars, the braking index is much lower than 3. The braking index obtained for a 1.4 $\textup{M}_\odot$ NS as a function of rotational frequency is shown in Fig. \ref{fig:br_fig}. The braking index varies very slowly and deviates a lot from the observations. Thus, an MDR model with a varying moment of inertia may not be sufficient to explain the observed pulsar braking indices. It is shown by \cite{br1, br2, br3}, and many others, that the braking index depends on various factors like the temporal evolution of the magnetic field strength and the inclination angle between the magnetic and the rotational axis. Although the braking indices of young pulsars are expected to be less than three, several estimations show that it can be greater than three as well \citep{Archibald_2016, parthasarathy1}. Even after decades of the discovery of pulsars, the braking index is not understood very well. The measurements of braking index presently do not provide strong constraints for the NS EOS. Future advancements in pulsar emission theory and its relation with the NS EOS will help in utilising such observations in constraining the NS EOS.

\section{Conclusions}

\label{conclusion}
In order to understand the interior structure of neutron stars, it is necessary that we constraint the equation of state with the maximum possible number of constraints from nuclear experiments and astronomical observations. In this work, we utilised several piecewise polytropic equations of state to check their compliance with astronomical observations. We demonstrated that a significant fraction of glitches cannot be explained with the fractional moment of inertia obtained for any of the equations of state. We modeled the neutron star crust as a solid spheroidal shell to calculate the fractional moment of inertia at high frequencies. We showed that the fractional moment of inertia increases monotonically with an increase in the rotation rate. Furthermore, we investigated the variation of braking index in the magnetic dipole radiation model, and showed that it cannot explain the observed data. The fractional moment of inertia and the braking index cannot be used to constrain the neutron star EOS but, with advancements in observational facilities, it may be possible in the near future.

\section*{acknowledgements}
This work is supported by the SPARK program of IIT Roorkee (India). J.S. acknowledges various discussions with V. B. Thapa. We thank P. Arumugam for his comments during the preparation of this manuscript. The authors especially thank the anonymous referee for their constructive comments which have improved the presentation of this article.

\bibliographystyle{raa}
\bibliography{main}

\end{document}